\documentclass[preprint, prb, showpacs,preprintnumbers,amsmath,amssymb,floats]{revtex4-1}
\usepackage{color,graphicx}
\usepackage{bm}
\usepackage{amsmath}
\usepackage{amssymb}
\usepackage{float}
\usepackage{epstopdf}
 \usepackage{color}
  \usepackage{ulem}

\newcommand{\be}{\begin{eqnarray}}
\newcommand{\ee}{\end{eqnarray}}

\newcommand{\FM}{YFe$_2$Al$_{10}$}
\begin{document}
\title{ Quantum critical singularities in two-dimensional metallic XY ferromagnets}
\date{\today}

\author{Chandra M. Varma}
\affiliation{Department of Physics, University of California, Riverside, CA. 92521}

\author{W. J. Gannon}
\affiliation{Department of Physics and Astronomy, Texas A\&M University, College Station, TX 77843}

\author{M. C. Aronson}
\affiliation{Department of Physics and Astronomy, Texas A\&M University, College Station, TX 77843}

\author{J. A. Rodriguez-Rivera}
\affiliation{NIST Center for Neutron Research, National Institute of Standards and Technology, Gaithersburg, MD 20899 USA}

\author{Y. Qiu}
\affiliation{NIST Center for Neutron Research, National Institute of Standards and Technology, Gaithersburg, MD 20899 USA}

\begin{abstract}
  An important problem in contemporary physics concerns quantum-critical fluctuations in metals. A scaling function for the momentum, frequency, temperature and magnetic field dependence of the correlation function near a 2D-ferromagnetic 
 quantum-critical point (QCP) is constructed, and its singularities are determined by comparing to the recent calculations 
  of the correlation functions of the dissipative quantum XY model (DQXY).  The calculations are motivated by the  measured  
 properties of the metallic compound YFe$_2$Al$_{10}$, which is a realization of the DQXY model in 2D. The frequency, temperature and 
 magnetic field dependence of the scaling function as well as the singularities measured in the experiments are given by the theory without 
 adjustable exponents. The same model is applicable to the superconductor-insulator transitions, classes of metallic AFM-QCPs,  and as 
 fluctuations of the loop-current ordered state in hole-doped cuprates.
The results presented here lend credence to the solution found for the 2D-DQXY model, and its applications in understanding quantum-critical 
properties of diverse systems.

\end{abstract}
\date\today
\maketitle

\section{Introduction}
{\FM} is nearly tetragonal, with a divergent uniform magnetic susceptibility at low temperatures with field applied in the a-c plane, but 
 a constant value at the same temperatures for fields applied along the b-axis \cite{AronsonPNAS2014}.  There is no observed anisotropy of 
the  susceptibility within the a-c plane.  These results suggest that  the metal is accidentally close to a ferromagnetic quantum critical point and that the relevant model for criticality is the 2D-XY model. The specific heat 
divided by temperature is logarithmic in temperature. We show here that the singularity in the susceptibility and the specific heat together and the singularity in the 
frequency/temperature dependence of the correlations \cite{Aronson2016} and their contrast with the momentum dependence are consistent with the recent 
solution of the 2D-DQXY model. 
 
 Classical 2D FM 
transitions of the Berezinskii, Kosterlitz-Thouless \cite{KT1973, Berezinskii} variety at finite T have been found in some insulating compounds in 
the past \cite{KT-FM}. {\FM} appears to be the first metallic compound to be very near a planar ferro-magnetic quantum-transition.

 \section{Response function of a 2D XY Model near quantum criticality}

 The 2D-dissipative quantum XY model describes the physics of interacting quantum rotors lying in a plane and includes dissipation due to transfer 
 of energy to other excitations. It is specified by the action given, for example, by Eq. (1) in Ref. {\onlinecite{ZhuChenCMV2015}}. Without 
 dissipation, the phase diagram and the correlation functions of the quantum XY model in 2D  belong to the classical 3D XY universality class. But in a 
 metal, the dissipation introduced by coupling of the fluctuations to corresponding incoherent fluctuations of the fermions, leads to a much 
 richer phase diagram \cite{ZhuChenCMV2015, Stiansen-PRB2012, footnote}.  A theory of the phase diagram and of the quantum-critical 
 fluctuations has been derived and tested by quantum Monte-Carlo calculations \cite{Aji-V-qcf1, ZhuChenCMV2015, HouCMVZhu2016}. The  
 fluctuations in such theories present a new paradigm in quantum critical phenomena. The conventional theories of quantum-critical phenomena 
 \cite{Moriya-book, Hertz} are based on anharmonic soft spin-fluctuations, which are extensions of the theory of classical dynamical critical 
 phenomena \cite{Hoh-Hal-RMP}, applicable to models of the Ginzburg-Landau-Wilson type.  In such theories, the frequency and momentum dependence 
 of the correlation function are always entangled and a finite dynamical exponent $z$ given by the dispersion of the spin-wave excitations in 
 the presence of dissipation relates the spatial and temporal correlations.
A quite different class of correlation functions are found for the 2D-DQXY model because the critical properties are determined not by spin-wave 
excitations but by topological excitations in space and time.

The 2D- DQXY model can be exactly transformed \cite{Aji-V-qcf1, Aji-V-qcf3} to a model of orthogonal topological charges, warps and vortices. 
Warps interact with each other in (imaginary) time and are essentially local in space while the vortices interact purely in space.  The correlation function of the order parameter $e^{i\theta({\bf r},\tau)}$ of the 2D-DQXY model  have 
been derived by quantum Monte-Carlo {\cite{ZhuChenCMV2015} which also checks their relation to the correlation functions of warps and vortices. 
The model transformed to interacting topological excitations has also been solved analytically {\cite{Hou-CMV-RG}}.
 The correlation function is found in an extensive region of parameters in which the proliferation of warps determines the criticality to be,
 \be
 \label{corrfn}
{\cal C}(r, \tau) \equiv \langle e^{i\theta({\bf r},\tau)}e^{-i\theta(0,0)} \rangle  \approx \chi_0 \log(r_0/r) e^{(-r/\xi_r)} \frac{1}{\tau} 
e^{-(\frac{\tau}{\xi_{\tau}})}.
\ee
The three especially note-worthy features of (\ref{corrfn}) are (i) it is separable in its $r$ and $\tau$ dependence, (ii) its thermal Fourier 
transform at criticality, when $\xi_{\tau} \to \infty$ has the   $\omega/T$ scaling \cite{Aji-V-qcf1}, introduced in critical phenomena in Ref. 
\cite{CMV-MFL} and termed "Planckian" \cite{Zaanen2004}, and (iii)
that \cite{ZhuChenCMV2015, Hou-CMV-RG}
\be
\label{xir-xit}
(\xi_r/a) = \log{(\xi_{\tau}/\tau_c)}.
\ee
This means that the dynamical critical exponent $z$ is effectively $\infty$. $\xi_{\tau}$ has an essential singularity as a function of the 
dimensionless dissipation parameter $\alpha$ but an algebraic singularity as a function of the dimensionless parameter $\tilde{K} \equiv 
\sqrt{KK_{\tau}}$. Here $K$ is the Josephson coupling and $K_{\tau}$ is the kinetic energy parameter in the quantum XY model. On 
the disordered side of the QCP, $\xi_{\tau}$ is given by,
\be
\label{xit}
\xi_{\tau}/\tau_c=   e^{\sqrt{c\alpha_c/(\alpha_c-\alpha)}}, ~at ~constant ~ {\tilde{K}}, ~~and = 
\Big(\frac{\tilde{K}_c}{\tilde{K}_c-\tilde{K}}\Big)^{\nu_{\tau}},~at ~ constant ~\alpha;~~ \nu_{\tau} \approx 1/2.
\ee
$c$ is a constant of $O(1)$ and $\tau_c$ a short-time cut-off.
If the transition, as expected is driven by $(\tilde{K}-\tilde{K}_c)$, the logarithmic dependence of the spatial correlation function may lead to 
a very short observed correlation length unless the sample is tuned to very small values of  $(\tilde{K}-\tilde{K}_c)$, and other effects, such as 
disorder do not change the asymptotic critical properties.

At criticality, i.e. for $\xi_{\tau}^{-1} = \xi_r^{-1} =0$, the {\it thermal Fourier transform} of the correlation function is
\be
{\cal C}(q,w,T) \propto  \frac{1}{q^2} \tanh{\frac{\omega}{2T}},
\ee
with a high frequency cut-off. For finite $\xi_{\tau}$ and $\xi_r$ the infra-red singularities are cut-off and their form is given in the Appendix in Ref.(\onlinecite{ZhuChenCMV2015}).

\section{Scaling for the Ferromagnetic quantum XY model \\ in a field}
The magnetic field $B_{\bot}$ in the plane couples linearly to the order parameter and serves as a cut-off to the quantum critical regime. To 
address the experimental results, we first present a scaling theory for the correlation function in a magnetic field, and connect the results to 
the calculated form, Eq. (\ref{corrfn}) derived at $B=0$.

A novelty is to derive a scaling form of the correlation function when the spatial correlations depend logarithmically on the temporal correlation 
length and neither may bear power-law relations to the control parameters. Consider the response of the 2D-XY ferromagnet with a uniform field $B$ 
in the easy plane at a temperature $T$ to a small applied time and space dependent field $h(r, t)$, also in the easy plane.  Follow the usual 
process of scaling for the correlation function on taking the derivative of logarithm of the partition function with respect to $h(r_1, \tau_1)$ 
and $h(r_2, \tau_2)$, $r = |r_1-r_2|, \tau = \tau_1-\tau_2$, and scale the space and time-metric together with the scaling operators in the action 
so as to keep the singular part of the partition function invariant. The space-metric is expanded by the correlation length  $\xi_r$ and the 
time-metric by $\xi_{\tau}$. The renormalization group eigenvalue for $B_{\bot}$ on scaling time is defined to be $z_b$.
 \be
\label{scaling}
{\cal C}(r, \tau, T, B_{\bot}) = \xi_r^{-2d} \xi_{\tau}^{-2}\xi_{\tau}^{2z_b} \chi\big(\frac{r}{\xi_r}, \frac{\tau}{\xi_{\tau}}, T \xi_{\tau}, 
B_{\bot} \xi_{\tau}^{z_b}\big).
\ee
 The $q = 0, \omega =0$ limit of the correlation function is found by integrating over $r$ and $\tau$. Divided by $T$, this gives the temperature 
 and magnetic field dependence of the static uniform susceptibility, Eq. (\ref{scaling2}). The integration over the space-variable brings a factor 
 $\xi_r^d$, as usual. At this point the special properties of the results in (\ref{corrfn}) may be used. Since the temporal correlation function 
 is  $\propto 1/\tau$ at criticality, integration over $\tau$ can produce at most only logarithmic corrections, which may be neglected to begin 
 with in comparison with the rest. Also, since $\xi_r \propto \log \xi_{\tau}$, the space dependent prefactors may also be neglected to 
 logarithmic accuracy. So we get
\be
\label{scaling2}
\chi(T,B_{\bot}) \equiv \frac{d M(T,B_{\bot})}{dB_{\bot}} &= &\frac{1}{T} <\cos^2(\theta)> = \frac{1}{T} C(q=0, \omega =0,T, B_{\bot}) \nonumber 
\\
& = & \frac{1}{T}\xi_{\tau}^{-2+2z_b} \chi \big(T\xi_{\tau}, B_{\bot}\xi_{\tau}^{z_b}\big).
\ee
On re-scaling $T\xi_{\tau} \to 1$ to express $\xi_{\tau}$ in terms of $T$,
one gets
\be
\label{res1}
\chi(T,B_{\bot}) \propto T^{(1-2z_b)}f_{1,\chi}\big(\frac{B_{\bot}}{T^{z_b}}\big),
\ee
or equivalently
\be
\label{res2}
\chi(T,B_{\bot}) \propto B_{\bot}^{(1-2z_b)/z_b} f_{2,\chi}(T/B_{\bot}^{(1/z_b)}).
\ee
On comparing (\ref{res1}) with the static susceptibility calculated from (\ref{corrfn}) and again neglecting logarithmic corrections, we find that 
the two are  mutually consistent only if $z_b=1$. Given the $1/T$ factor in (\ref{scaling2}), the correlation function has an exponent $0$ which 
is consistent with having logarithmic corrections. Scaling cannot give the logarithmic corrections, which turn out to be important in relation to experiments, as seen below. We therefore explicitly calculate the magnetic susceptibility by the Monte-Carlo technique using the procedure of 
Refs. \onlinecite{ZhuChenCMV2015} for the dissipative quantum XY model.

 \subsection{Monte-Carlo Calculations:}

 The uniform magnetic susceptibility per unit-cell is
 \be
 \chi(T) \equiv \frac{1}{N^2} \sum^{N^2}_i \int_0^{\beta} d\tau <M(i,\tau)M(i,0)>;~~ M(i,\tau) = \cos(\theta_i)(\tau),
 \ee
 where $N^2$ is the number of unit-cells on a lattice labelled by $i$.
This is converted to a form suitable for quantum Montecarlo calculations on a discrete space $N\times N$ and imaginary time one-dimensional 
lattice $\tau_n$ of $N_{\tau}$ cells,
 \be
\chi (T) =\frac{1}{N^2 N_{\tau}} \sum^{N^2,N_{\tau}}_{i,n} <\cos(\theta_i,n)\cos(\theta_i,0) >.
\ee
The calculation is entirely as in the calculation of the action susceptibility, Eq. (10), of Ref.~\onlinecite{ZhuChenCMV2015}. The discretization and calculation procedure is also fully described there in Sec. II-C. 
$\tau_n = n \delta\tau = n \beta/N_{\tau}$. $\delta \tau = \tau_c$ is the ultra-violet (short) time cut-off. The temperature in the calculation is 
controlled by $N_{\tau}^{-1}$. This has been calculated on a $N^2 =50 \times 50$ lattice and with $N_{\tau}$ ranging from 20 to 200. With an 
upper-cutoff $\tau_c^{-1} = 160 K$, this effectively gives results at discrete temperatures from 20 to 0.8 K. The results for $\chi(T)$ are 
given in Fig. (\ref{Fig:Chi}).

The black crosses in Fig. (\ref{Fig:Chi}) are the result, and they compare favorably to the measured uniform susceptibility $\chi$, also shown in 
Fig.~\ref{Fig:Chi}.  Motivated by the discussion above, we look for logarithmic factors multiplying $T^{-1}$. We find that the calculated 
susceptibility fits $T^{-1} \log^2(T/\omega_c)$, where $\omega_c = 1/\tau_c$ is the high energy cut-off given in Eq. (\ref{xit}). We also show the 
experimentally derived function $T^{-1.4}$, which mimics  $\frac{1}{T} \big(\log T \tau_c \big)^{2}$ very well over the range of experimental 
temperatures, with $\tau^{-1}_c = 160~{\rm K}$.  The previously reported scaling analysis\cite{AronsonPNAS2014} is purely phenomenological, with 
two critical  exponents that are determined by the experiments and with a spatial correlation length, discussed below, which is in qualitative conflict with experiments. In contrast, the logarithmic corrections found here leave no parameter in the 
theory undetermined.


\begin{figure}[tbh]
\includegraphics[width=0.6\columnwidth]{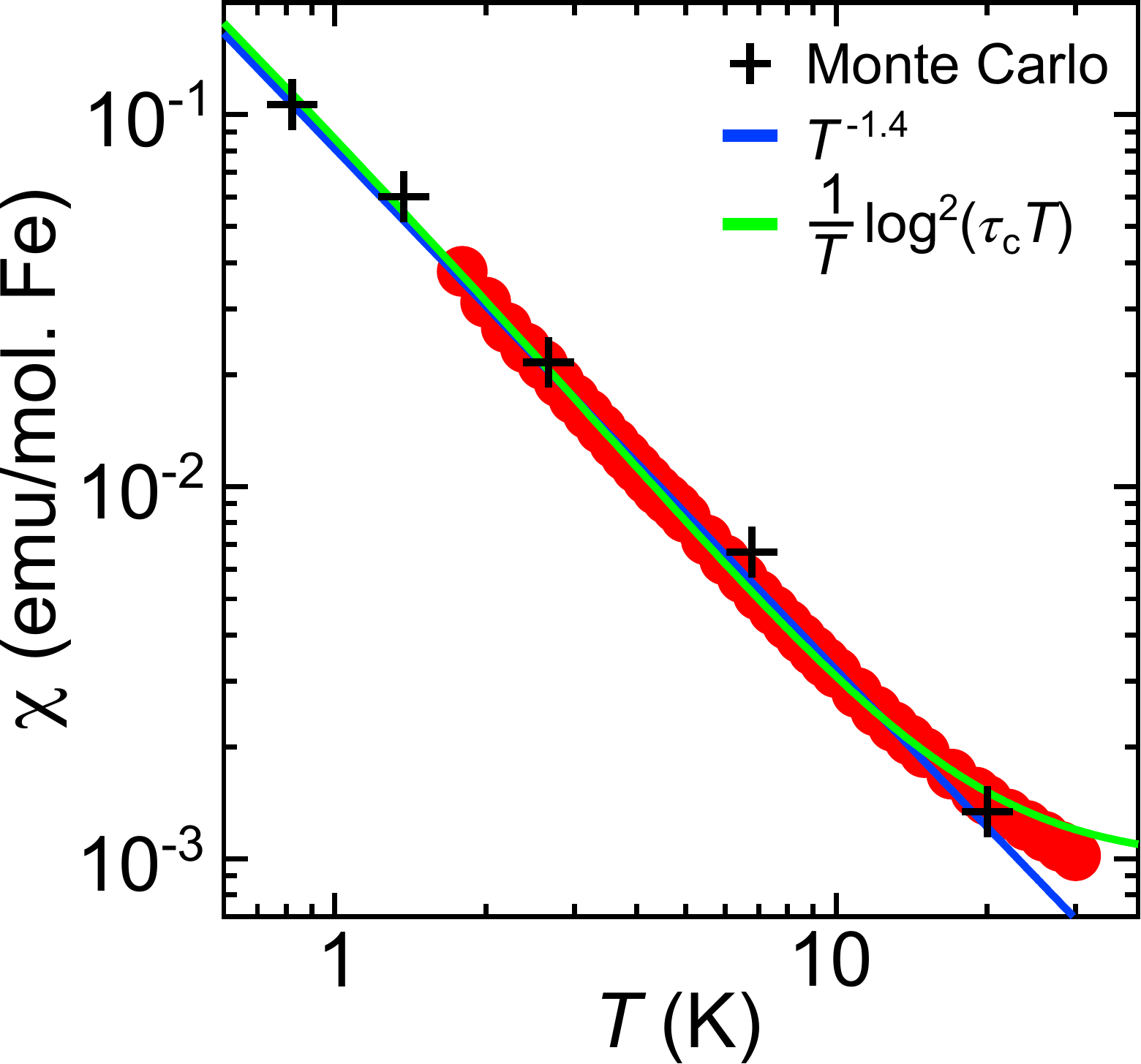}
\caption{Comparison between the uniform magnetic susceptibility $\chi=M/B$ of YFe$_2$Al$_{10}$ measured in a field of  $0.05~\rm T$ in the 
critical $a-c$ plane (red circles) \cite{AronsonPNAS2014} to fits to $\propto T^{-1.4}$ (blue line) and the form calculated from the correlation 
functions of the dissipative 2D-quantum XY Model in this paper with $\tau_c^{-1}$ = 160 K, (green line) which is the approximate scale below which 
the divergent form appears in the experiments. Also shown are direct calculation of $\chi$ by Monte-Carlo method (black crosses), with temperature 
and susceptibility scaled to the experimental data. Fits are performed for $T<20~\rm K$ with $\tau^{-1}_c=160~K$ as a fitted parameter in the 
theory for the 2DXY model.}
\label{Fig:Chi}
\end{figure}

 The experimental results \cite{AronsonPNAS2014} for the scaling of $M(B,T)$ in $YFe_{2}Al_{10}$, previously fitted \cite{AronsonPNAS2014} to the 
 scaling expression
\be
\label{scaleM}
-\big(d(M/B_{\bot})/dT\big) B_{\bot}^{1.4} \propto F(T/B_{\bot}^{(1-0.4)}),
\ee
are compared to the result
\be
\label{scaleM2}
-\big(d(M/B_{\bot})/dT\big) \frac{B_{\bot}}{\log^2(B\tau_c)} \propto f_M\Big(\frac{T}{B_{\bot} \log^2(B\tau_c)}\Big).
\ee
\
in Fig. 2.

\begin{figure}[tbh]
\includegraphics[width=1.0\columnwidth]{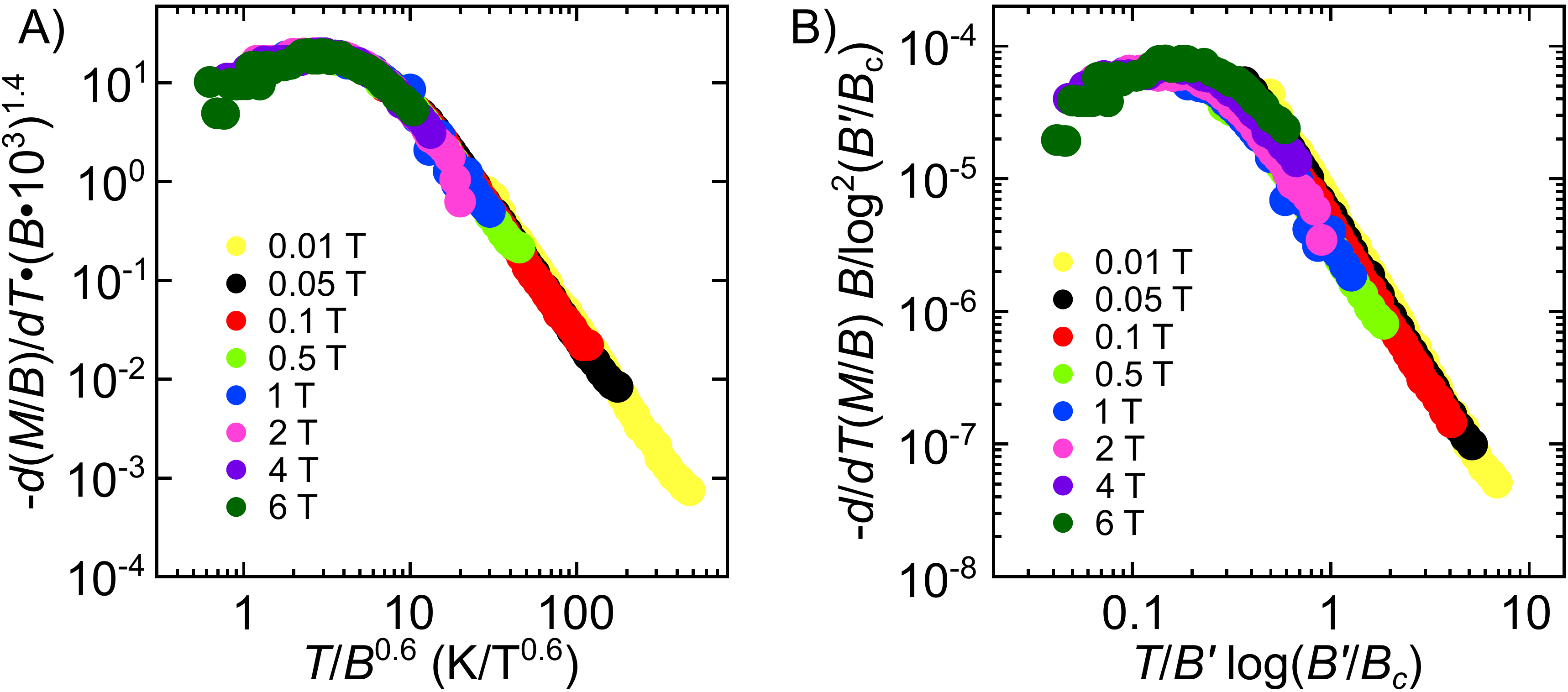}
\caption{(A) The scaling of the uniform magnetization divided by field  $M/B$ as a function of temperature $T$  measured at different fixed fields 
$B$  indicated in the figure to Eqn.~\ref{scaleM}, as shown in~\cite{AronsonPNAS2014}.  Measurements were made for temperatures $1.8\leq T \leq 
30~\rm K$ with the field in the critical $a-c$ plane. (B)  Same data as (A), which can be scaled using Eqn.~\ref{scaleM2}, with a substitution of 
the external field $B$ with $B'=B+0.07~\rm T$ and $B_c=100~\rm T$.  Colors indicating different magnetic fields are the same as (A).}
\label{Fig:ChiScale}
\end{figure}

Fig.~2 shows that Eqn.~\ref{scaleM2} gives an acceptable scaling collapse, with a tiny off-set of the field $B\rightarrow B'$, $B'=B+0.07~\rm T$ 
and $B_c=100~\rm T$, the latter in reasonable agreement with the value of $\tau^{-1}_c=160~\rm K$, using $g\mu_BB_c=\hbar/\tau_c$ with the Land\'e 
g-factor taken to be 2 and $\mu_B$ the Bohr magneton. We do not know the origin of the small off-set of 0.07 T required to best fit the data which spans the range up to 
6 T; it may be due to impurities in the sample.}\

\subsection{Scaling of the Free-energy:}

The scaling for the free-energy per unit volume may be considered similarly
\be
\label{res4}
f(T,B) \propto T \xi_r^{-d} \xi_{\tau}^{-1} \Phi\big(T\xi_{\tau}, B_{\bot}\xi_{\tau}^{z_b}\big)
\ee
This gives, using the same results as for the calculation of magnetization, that
\be
\label{resf}
f(T,B) \propto T^2 \ln^2(T\tau_c) \tilde{\Phi}\Big(\frac{B_{\bot}}{T/\log^2(T\tau_c)}\Big).
\ee
With (\ref{resf}), the results for $M(T,B_{\bot})$ and $\chi(T,B_{\bot})$ derived above from the correlation functions follow to logarithmic 
accuracy. The specific heat divided by $T$ at constant $B_{\bot}$ has in addition to a constant and a $\log(T)$  term a $\log^2(T)$ term with a 
coefficient that is 1/3 of the logarithmic term. The specific heat as a function of magnetic field $B_{\bot}$, similarly follows. Note the factor 
$T$ in (\ref{res4}). This is un-important for classical transitions, where it is replaced near criticality by $T_c$ but essential to keep for a 
transition with $T\to 0$\\

\subsection{Dynamics}

Consider now the extension of the correlation function, Eq. (\ref{scaling}) to obtain the frequency and momentum dependent magnetic response function. In the absence of the 
detailed Monte-Carlo calculations of the correlation function in a magnetic field, one may guess on grounds given below that the magnetic response 
function has the approximate scaling form,
\be
\label{chiqw}
{\chi}"(q,\omega, T, B_{\bot}) \propto \frac{\log(\sqrt{\omega^2 + (2k_B T)^2}\tau_c)}{\sqrt{\omega^2 + (2k_B T)^2}} f_{\chi}\Big(\frac{\omega}{T}, qa 
\log(\xi_{\tau}/\tau_c), \frac{B_{\bot}}{(T/\log(\sqrt{\omega^2 + (2k_B T)^2}\tau_c)}\Big).
\ee
This follows the form of the derived correlation function (\ref{corrfn}) except for the modifications necessary due to the scaling corrections due 
to $B_{\bot}$. The logarithmic term and its argument have been chosen so that it reproduces the temperature dependence of the calculated uniform 
magnetic susceptibility, derived by using the Kramers-Kronig relation between the imaginary part $\chi"(q,\omega, T, B_{\bot})$ and the real part at 
$\omega =0$, as well as the magnetic field dependence of the magnetization derived above.

Eq. (\ref{chiqw}) may be put in various other forms as desired. It follows that for finite $B_{\bot}$, this divergence is cut-off.  It is 
predicted that together with $\omega/2T$ scaling of the form calculated in microscopic theory to be of the form $\tanh(\omega/2T) $, with a 
cut-off at $\omega_c = \tau_c^{-1}$, there should be singular pre-factors. This has been tested by inelastic neutron scattering as described in 
Ref. \cite{Aronson2016}, where it is found that a cutoff energy dependence of the form $\left(\omega^2+\left(\pi 
T\right)^2\right)^{-1.4/2}\tanh\left(\omega /T\right)$ describes the data reasonably well. The data  can be equally fitted by Eq.~(\ref{chiqw}).

\begin{figure}[tbh]
\includegraphics[width=0.6\columnwidth]{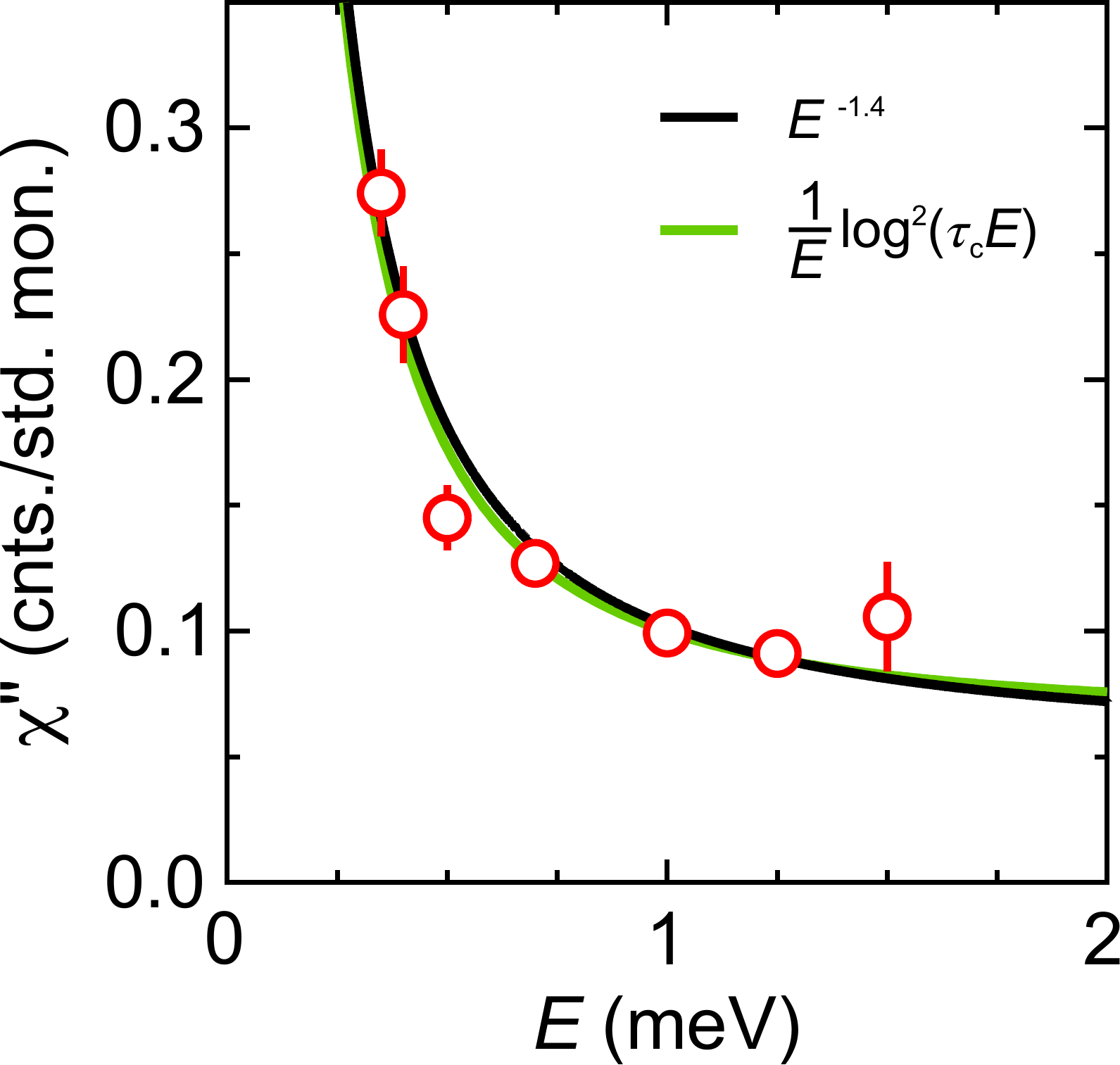}
\caption{The dynamical susceptibility $\chi''$ at $T=0.1 K$ as a function of energy obtained by inelastic neutron scattering using the MACS spectrometer at the National Institute For Standards and Technology~\cite{Rodriguez2008}.  Data (red circles) were measured at a constant fixed energy over a large area of $q$-space, corrected for the Fe$^{2+}$ form factor, then integrated over four Brillouin zones that are out of the critical ($a-c$) plane (along $q_K$, parallel to the crystal $b$-axis) and over one Brillouin zone within the critical plane (along $q_L$, parallel to the crystal $c$-axis)~\cite{Aronson2016}.  Measurements were made with a small bias field of 0.025 T along the crystal $a$-axis to suppress superconductivity in the aluminum sample holder.  Fits are made to the forms given by Eq. (\ref{chiqw}) for $E = \hbar \omega >> k_BT$, so that $\chi'' \sim E^{-1.4}$ (black line), and $\left(1/E\right)\mathrm{log}^2\left(E~\tau_c\right)$ (green line), fixing ($\tau^{-1}_c=14~\mathrm{meV} \approx 160 ~K$).  Error bars on data points represent one standard deviation.}
\label{Fig:Chiimag}
\end{figure}

A comparison of Eq. (\ref{chiqw}) presented here to the energy dependence of the measured dynamical susceptibility is shown in Fig.~\ref{Fig:Chiimag} for $\omega >> T$.   The correspondence between temperature and energy revealed by a previous scaling analysis~\cite{AronsonPNAS2014} and the Kramers-Kroning relation~\cite{Aronson2016} suggests that the momentum-integrated dynamical susceptibility $\chi''$ is a function of $\left(1/\omega\right)\mathrm{log}^2\left(\omega \tau_c\right)$ with a high energy cutoff $\omega_c$.   Fixing  $\tau^{-1}_c=14~\mathrm{meV}$ ($\approx 160~\mathrm{K}$) gives the fit shown in Fig.~\ref{Fig:Chiimag}, nearly indistinguishable from the phenomenological $E^{-1.4}$ power law behavior used previously~\cite{Aronson2016}. The correspondence of Eq. (\ref{chiqw}) to the change in the dynamics in a magnetic field may be seen in that paper.

On considering the q-dependence, one encounters an interesting discrepancy in relation to the experiments. The sample is, not surprisingly, not 
exactly at criticality. The dynamical measurements, both through neutron scattering and more directly through the muon spin-relaxation rate 
\cite{muons1210}  suggest a low temperature cut-off in the experiments  of about 1 K.  So $\xi_{\tau}^{-1} \approx 1 K$.
 In experiments not exactly at criticality, $T$ and $B_{\bot}$ should be replaced approximately by $\sqrt{T^2 + 
 \xi_{\tau}^{-2}}$ and $\sqrt{B_{\bot}^2 + B_x^2}$ with
$(g\mu_B B_x)^2 <S^2> \approx \xi_{\tau}^{-2}$, where $(g\mu_B)^2<S^2>$ is a measure of the mean-square
magnetic moment in the fluctuations.
Using $\tau_c^{-1} \approx 100 K$, the corresponding cut-off in the spatial correlation length $\xi_r$ may be estimated using Eqs. (\ref{xir-xit}) 
to be about 4 lattice constants. But the spatial correlation length in neutron scattering experiments is only about a lattice constant, although 
independent of temperature in accord with the theory. A possible explanation \cite{CMV-crossover} of such extreme spatial locality while 
scale-invariant behavior is observed with long temporal correlation length $\xi_{\tau}/\tau_c$ of $O(10^2)$ may lie in the crossover due to 
disorder in quantum-critical problems with large dynamical critical exponent $z$. This matter can be tested by further experiments in samples 
closer to criticality. Tuning closer to criticality may be difficult since using the second of (\ref{xit}), which is the more likely applicable, 
$\xi_{\tau}/\tau_c \approx 10^2$ implies that already $(1-\tilde{K}/\tilde{K}_c) \approx 10^{-3}$.

\section{Concluding Remarks}
These results test the theory of the 2+1 D - XY model in considerable detail.  In particular, the success of the results in explaining the 
singularities in the properties associated with the free energy depends on the novel results of the theory that the  correlation function is the 
product of a function in space and a function in time, and that the spatial correlations vary logarithmically as the temporal correlations. The 
result that at criticality the time-dependence is proportional to $1/\tau$, i.e has the Planckian scaling $\omega/T$, has also been crucial. As 
may easily be seen, these results cannot be obtained by simply putting the dynamical exponent $z \to \infty$ in the conventional dynamical 
critical theory. Further tests of the theory require samples in which the distance to quantum-criticality can be systematically changed, for 
example, by applying pressure, thereby observing a longer spatial correlation length varying logarithmically as the distance to the critical 
point.

{\it Acknowledgements}: CMV acknowledges with pleasure discussions with Joerg Schmalian and Alexei Tsvelik.  Special thanks are due to Changtao 
Hou and Lijun Zhu who wrote the Monte-Carlo routines used for the results shown in Fig. (1).  Part of this research was conducted at Brookhaven 
National Laboratory, where  W. J. G and M. C. A  were supported under the auspices of the US Department of Energy, Office of Basic Energy 
Sciences, under contract DE-AC02-98CH1886.  Access to MACS was provided by the Center for High Resolution Neutron Scattering, a partnership between the National Institute of Standards and Technology and the National Science Foundation under Agreement No. DMR-1508249.


\bibliography{CorrectReferences_WJG}

\begin{thebibliography}{20}
\bibitem [{foo()}]{footnote}%
  \BibitemOpen
  \href@noop {} {}\bibinfo {note} {A Caldeira-Leggett form of dissipation is
  used in Refs. 6, 9, and 10. In the present context, this can be shown to come
  from the coupling $\nabla \theta$, which represents the collective
  ferromagnetic spin-current, to the spin-current of the incoherent
  fermions.}\BibitemShut {Stop}%
\end{thebibliography}
\end{document}